\documentclass[prd, twocolumn, aastex]{revtex4-1}
\pdfoutput=1 %for arXiv submission
\usepackage{amsmath,amstext}
\usepackage[T1]{fontenc}
\usepackage{apjfonts} 
\usepackage{graphicx}
\usepackage{natbib}
\usepackage{xcolor}
\usepackage{hyperref}

\usepackage{aasjournal}

\begin{document}

\title{Forecasting Cosmological Constraints from the Weak Lensing Magnification of Type Ia Supernovae Measured by the Nancy Grace Roman Space Telescope}

\author{Zhongxu Zhai$^*$}

\author{Yun Wang}

\affiliation{IPAC, California Institute of Technology, Mail Code 314-6, 1200 E. California Blvd., Pasadena, CA 91125}

\author{Dan Scolnic}

\affiliation{Duke University, Department of Physics, Durham, NC 27708, USA}

\email{zhai@ipac.caltech.edu}
% Enter the current year, for the copyright statements etc.

\begin{abstract}

The weak lensing magnification of Type Ia Supernovae (SNe Ia) is sensitive to the clustering of matter, and provides an independent cosmological probe complementary to SN Ia distance measurements. The Nancy Grace Roman Space Telescope is uniquely sensitive to this measurement as it can discover high redshift SNe Ia and measure them with high precision.
We present a methodology for reconstructing the probability distribution of the weak lensing magnification $\mu$ of SNe Ia, $p(\mu)$, from observational data, and using it to constrain cosmological parameters. 
We find that the reconstructed $p(\mu)$ can be fitted accurately by a stretched Gaussian distribution, and used to measure the variance of $\mu$, $\xi_\mu$, which can be compared to theoretical predictions in a likelihood analysis. Applying our methodology to a set of realistically simulated SNe Ia expected from the Roman Space Telescope, we find that using the weak lensing magnification of the SNe Ia constrains a combination of matter density $\Omega_m$ and matter clustering amplitude $\sigma_8$. SN Ia distances alone lead to a better than 1\% measurement of $\Omega_m$. The combination of SN Ia weak lensing magnification and distance measurements result in a $\sim$ 10\% measurement on $\sigma_8$. The SNe Ia from Roman will be powerful in constraining the cosmological model.

\end{abstract}


\maketitle

\section{Introduction}

As a key cosmological probe, Type Ia supernovae (SNe Ia) provided the first direct evidence of the acceleration of the expansion of the Universe \citep{Riess_1998, Perlmutter_1999}, a.k.a. "dark energy", via their observed luminosity distance-redshift relation as calibrated "standard candles". This method falls into the category of geometrical probes, i.e. sensitive to the background expansion of the universe. This type of probe also includes the ``standard ruler" which can measure the cosmic distance scales through Baryon Acoustic Oscillations (e.g.\cite{ Blake_2003, Eisenstein_2005, Anderson_2014, Alam_2016}). Various surveys over the past several decades have obtained data for thousands of SNe Ia events \citep{Riess_1999, Riess_2004, Astier_2006, Miknaitis_2007, Conley_2011, Frieman_2008, Suzuki_2012, Rest_2014, Graur_2014}. The latest compilation of the SN Ia dataset can measure the dark energy equation of state $w$ to 3-4\%, when combined with constraints from the CMB \citep{Amanullah_2010, Betoule_2014, Scolnic_2018}. 

SN Ia observations also contain information beyond the measurement of cosmic expansion. This possibility has been investigated by several authors in the recent years, through the weak gravitational lensing effect. Since the matter in the universe is not distributed with perfect homogeneity, the light received by the observer from a distant object is bent along the line of sight. Therefore the observed brightness of SNe Ia can have a distribution different from the intrinsic brightness, i.e. magnification. Early investigation such as \cite{Wang_2005_wl} reports a detection of this weak lensing magnification effect in a high-redshift SNe Ia sample. Recent analysis in \cite{Zhai_2019} with the latest Pantheon sample reports a $2\sigma$ signal. These studies are based on the assumption that the weak lensing effect can be expressed in terms of a probability function of the magnification \cite{Valageas_2000a, Wang_2002, Vale_2003, Takahashi_2011}, and the resulting distribution of the observed SNe Ia brightness is a convolution of this magnification distribution and the intrinsic brightness distribution. The limited size of the current SNe Ia data set severely hampers the reconstruction of the weak lensing magnification signal to constrain cosmology. 
In this paper, we employ realistic simulations of SNe Ia from the Nancy Grace Roman Space Telescope to explore the implications of SN Ia magnification distributions for probing cosmology .

The weak lensing signature is sub-dominant compared with the intrinsic brightness distribution of SNe Ia, but its amplitude grows with increasing redshift. Future surveys, such as those planned for Roman Space Telescope \citep{Spergel_2015} and Rubin Observatory \cite{LSST-sciece-book}, will collect high quality data of at least tens of thousands of high-redshift SNe Ia.
Rubin will discover SNe out to $z\sim 1.2$, whereas Roman will discover SNe to $z\sim 2.5$, which makes lensing measurement much easier. SNe Ia from Roman will enable a detailed investigation of the weak lensing magnification effect. 
This will enhance the power of SNe Ia as a cosmological probe beyond that of a geometrical probe, by providing constraints based on the growth of large scale structure in the universe, which will lead to improvements on the constraints on the dark energy models and modified gravity theories. This approach has been visited in the literature, e.g. \cite{Dodelson_2006, Marra_2013, Quartin_2014, Castro_2016, Castro_2018, Macaulay_2017, Scovacricchi_2017, Zumalacarregui_2018} and references therein. This includes utilizing the observed SN Ia magnitude residuals in a Hubble diagram \cite{Dodelson_2006, Macaulay_2017}, the ``MeMo" likelihood methodology to characterize the non-Gaussian distribution of the SNe Ia magnitude residuals \cite{Marra_2013, Quartin_2014}, the magnitude angular correlation function \cite{Scovacricchi_2017},  the impact on the neutrino property constraints \cite{Hada_2016, Hada_2018} and so on. We refer the readers to the above references for more details. 

In our earlier work \cite{Zhai_2019}, we developed a method to extract the distribution function of the lensing magnification $p(\mu)$ from the latest SN Ia data compilation. We extend that earlier analysis by utilizing a realistically simulated SN Ia data set from Roman Space Telescope in this work. We adopt the method from \cite{Zhai_2019} to model the underlying magnification distribution function $p(\mu)$ from observational data, and compress the results into physically intuitive quantities to enable a likelihood analysis to extract cosmological constraints. This enhances the cosmological constraints from SNe Ia beyond that of a geometrical probe only, and helps break the degeneracy between cosmological parameters. This will eventually help shed light on the apparent tensions between different observations at present \cite{Planck_2018_6, Kohlinger_2017}.

Our paper is organized as follows. In Section 2, we present the modeling of the weak lensing signature in SNe Ia observations. In Section 3, we introduce the simulated data, the $p(\mu)$ reconstruction method and its cosmological implications. Section 4 presents our analysis results.
We conclude in Section 5 with a summary and discussions.

\section{Weak lensing signature of SNe Ia }\label{sec:signature}

The weak lensing magnification of SNe Ia has been discussed extensively in literature, e.g. \cite{Bernardeau_1997, Kaiser_1998, Valageas_2000a, Valageas_2000b, Wang_2002}. Here we summarize the key results relevant for the analysis from \citep{Wang_2005_wl}.

The observed flux from a SN Ia can be written as
\begin{equation}
    f=\mu L_{\text{int}},
\end{equation}
where $L_{\text{int}}$ is the intrinsic brightness of the SN Ia, and $\mu$ is the magnification due to weak lensing, which can be modeled by a universal probability distribution function based on the measured matter power spectrum \citep{Wang_2002}. The two variables $L_{\text{int}}$ and $\mu$ are assumed to be statistically independent, therefore the distribution of their product $f$ can be modeled explicitly with the probability distribution function (PDF) of the variables. The resulting distribution can be written as 
\begin{equation}\label{eq:pdf_f}
    p(f)=\int_{0}^{L_{\text{int}}^{\text{max}}}\frac{dL_{\text{int}}}{L_{\text{int}}}g(L_{\text{int}})p\left(\frac{f}{L_{\text{int}}}\right),
\end{equation}
where $p(f/L_{\text{int}})= p(\mu)$ is the PDF of the magnification $\mu$, and $g(L_{\text{int}})$ is the PDF of the intrinsic brightness of SNe Ia. The integral is from 0 to an upper limit  $L_{\text{int}}^{\text{max}}=f/\mu_{\text{min}}$, due to the requirement $\mu=f/L_{\text{int}}\geq\mu_{\text{min}}$, where $\mu_{\text{min}}$ is the minimum value of the magnification due to lensing and can be computed for a given cosmological model. We adopt the same assumption as in \cite{Wang_2005_wl, Zhai_2019} that $g(L_{\text{int}})$ is a Gaussian distribution with dispersion $\sigma$. The value of $\sigma$ can be well estimated with a large sample of SNe Ia at low redshifts where the weak lensing effect is negligible.

The cosmological model dependence of weak lensing magnification is encoded in $p(\mu)$, and we note that there are multiple ways for its computation, including both analytic method and numerical methods based on N-body simulations \cite{Valageas_2000a, Barber_2000, Premadi_2001, Vale_2003, Wambsganss_1997, Yoo_2008, Takahashi_2011}. Here we assume $p(\mu)$ can be modeled by the universal probability distribution function (UPDF) \cite{Wang_1999, Wang_2005_wl}
\begin{equation} \label{eq:P_eta}
    p(\eta)=\frac{1}{1+\eta^2}\exp\left[ -\left(\frac{\eta-\eta_{\text{peak}}}{\omega\eta^{q}}\right)^{2} \right],
\end{equation}
where 
\begin{equation} \label{eq:P_eta2}
    \eta=1+\frac{\mu-1}{|\mu_{\text{min}}-1|}.
\end{equation}
The parameters in this formula, $\{\eta_{\text{peak}}, \omega, q\}$, are functions of the variance of $\eta$, $\xi_{\eta}$, which absorbs all the cosmological dependence. Together with $\mu_{\text{min}}$, the minimum of the magnification, the parameter set $\{\mu_{\text{min}}, \eta_{\text{peak}}, \omega, q\}$ is able to completely determine the $p(\mu)$ model and therefore the theoretical prediction for the distribution of observed brightness of SNe Ia including the effect of weak lensing magnification. 

In principle, $p(\mu)$ can be used as the observable in comparing predictions with observations. For simplification, here we only use its moments in explicit comparison of model with data. For an arbitrary cosmological model, one can compute $\xi_{\eta}$ as \citep{Valageas_2000a}
\begin{equation}\label{eq:xi_mu}
    \xi_{\mu}=\int_{0}^{\chi_{s}}d\chi w^{2}(\chi, \chi_{s})\,I_{\mu}(\chi),
\end{equation}
with
\begin{eqnarray}
    &&I_{\mu}=\pi\int_{0}^{\infty}\frac{dk}{k}\frac{\Delta^2(k,z)}{k}W^2(Dk\theta_{0}), \nonumber \\
    && \Delta^2(k,z)=4\pi k^3P_{m}(k,z), \quad W(Dk\theta_{0})=\frac{2J_{1}(Dk\theta_{0})}{Dk\theta_{0}} 
    \label{eq:Imu}
\end{eqnarray}
where $P_{m}(k, z)$ is the matter power spectrum at redshift $z$ with wavenumber $k$, $\theta_{0}$ is the smoothing angle \citep{Valageas_2000b}, and $J_{1}$ is the Bessel function of order 1. The other quantities depending on the distance measure in the universe, and can be calculated as follows
\begin{eqnarray}
    && w(\chi, \chi_{s}) = \frac{H_{0}^2}{c^{2}}\frac{D(\chi)D(\chi_{s}-\chi)}{D(\chi_{s})}(1+z)  \nonumber \\
    && D(\chi) = \frac{cH_{0}^{-1}}{\sqrt{|\Omega_{k}|}}\text{sinn}\left(\sqrt{|\Omega_{k}|}\,\chi\right), \nonumber \\
    && \chi = \int_{0}^{z}\frac{cH_{0}^{-1}dz'}{E(z')}, \nonumber\\
    && E(z) = \frac{H(z)}{H_0}
\end{eqnarray}
where ``sinn" is defined as sinh if $\Omega_{k}>0$, sin if $\Omega_{k}<0$. If $\Omega_{k}=0$, both sinn and $\Omega_{k}$ disappear. Higher order moments of $\eta$ can provide additional information, but their accurate calculation requires calibration from simulations \cite{Valageas_2000b}, thus we leave these for future work and focus on the variance $\xi_{\eta}$ in this paper. 

The smoothing angle $\theta_{0}$ is a nuisance parameter from the window function for computing the variance of the weak lensing magnification, and degenerate with the amplitude of the matter power spectrum (see Eq.[\ref{eq:Imu}]). This leads to the degeneracy between cosmological parameters and $\theta_0$. Fortunately, this degeneracy can be removed by measuring $\theta_0$ from cosmological ray-shooting simulations, see discussion in Sec.\ref{sec:lensing_constraints}.

\section{Simulation of SNe Ia}

\subsection{Modeling SN Ia Systematic Effects}
\label{sec:sys}

To build realistic simulations of the Roman SN survey, we follow the strategy and design explained in \cite{Hounsell18}.  Here we use their 'All-z' survey, which has a shallow, medium and deep tier, where each uses four filters for observations every 5 days.  The four filters are $RZYJ$, $RZYJ$ and $YJHF$ for the three tiers respectively and the areas covered for each are $48.82, 19.75, 8.87$ square degrees.  In total, simulations predict that up to $14,000$ SNe Ia may be discovered up to $z\sim3$. 

To create the simulations, we use the \textit{SNANA} simulation package \citep{Kessler09} which produces high-fidelity catalogs of the expected photometric light-curves of the SNe.  The simulations are based on a description of the observatory (filter properties, zero-points, sky noise, PSF sizes), the survey (cadence, exposure times, and detection/selection requirements) and a description of the physical universe (SN rates, the SALT2 spectral model from \cite{Betoule_2014}, cosmological parameters).  The simulations include a model of the intrinsic scatter of SNe Ia based on \cite{Guy2010}, which can be described as 75\% achromatic variation and 25\% chromatic variation and parameters for the color and stretch population derived in \cite{Scolnic16}.  \textit{SNANA} can incorporate lensing models within the simulations, however here they are added posteriori to understand specific effects.  Therefore, it is assumed in this analysis that lensing does not contribute strongly to the impact of SN selection relative to typical SN variation.

\begin{figure}%[htbp]
\begin{center}
\includegraphics[width=8cm]{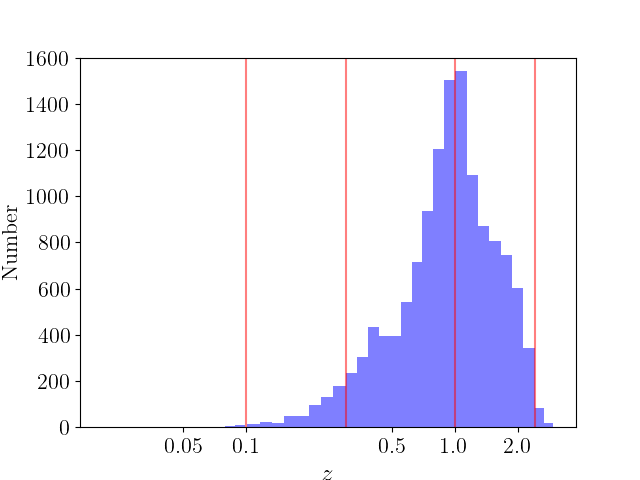}
\caption{The redshift distribution of the simulated SNe Ia data sample. The vertical red lines mark the redshift cuts used in this analysis. $SNe~Ia$ with $z<0.1$ are impacted by the peculiar velocity significantly, the data within $0.1<z<0.3$ are used to anchor the mean flux and model the distribution of the intrinsic brightness, the weak lensing signal for cosmological analysis is limited to $1.0<z<2.4$.}
\label{fig:sim_z}
\end{center}
\end{figure}

To measure distances from the simulated light-curves, we again use the SALT2 model to fit the light-curves and then follow \cite{Marriner11} to determine nuisance parameters and convert the light-curve parameters to distance modulus values.  Following \cite{Hounsell18}, we apply conventional light-curve quality cuts to ensure accurate and precise distances.  The redshift distribution of the simulated SNe is shown in Fig. \ref{fig:sim_z}, along with the redshift cuts we use for the analysis in this paper.

In order to validate our analysis methodology, we have created a "No SYS" companion data set of SNe Ia as follows, as a baseline for comparing with the realistic "SYS" data set described above.
We first estimated the SN Ia intrinsic flux distribution from the low-$z$ subsample, and found that it can be well described by a Gaussian model with standard deviation $\sigma=0.1$ (see Fig.\ref{fig:lowz}). We then created the "No SYS" simulated data set containing SNe Ia at the same redshifts as the "SYS" data set.
For each SN Ia in the "No SYS" set, we take its distance modulus to be randomly drawn from a Gaussian distribution with standard deviation $\sigma=0.1$, and mean given by the prediction from the input cosmological model at that redshift.
We add weak lensing signal to both the "SYS" and "No SYS" data sets, and compare the cosmological constraints derived, see Table \ref{tab:terms}.

\subsection{Modeling Weak Lensing Magnification of SNe Ia}

\begin{figure}%[htbp]
\begin{center}
\includegraphics[width=8cm]{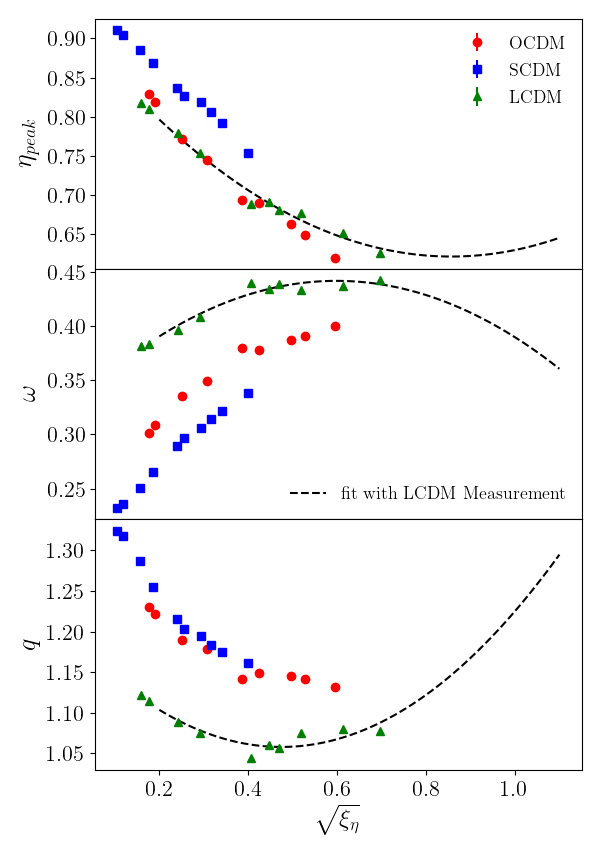}
\caption{The dependence of the parameters to model $p(\eta)$ as in Eq.(\ref{eq:P_eta}) on the value of $\eta$, for different cosmological models.}
\label{fig:pmu_parameter_fit}
\end{center}
\end{figure}

The weak lensing model for the SNe Ia brightness is described by the probability function $p(\mu)$ or $p(\eta)$. The previous works in \cite{Wang_1999, Wang_2002} assume $p(\eta)$ is universal, i.e. it can be well approximated by the stretched Gaussian distribution (Eq.[\ref{eq:P_eta}]) and the values of the parameters depend only on $\xi_{\eta}$. The result is found to be in agreement with numerical ray-tracing simulations. In this section, we revisit the universality of $p(\eta)$ by comparing the model Eq.(\ref{eq:P_eta}) with the measurements from simulations. The ray-tracing simulations we use were carried out following the methodology from \cite{Barber_2000}, and provided by Andrew Barber (private communication). The cosmological parameters and simulation details are summarized in Table 1 of \cite{Barber_2000}. 
They studied four cosmological models:
SCDM ($\Omega_{m}=1.0, \sigma_{8}=0.64, \Gamma=0.5$), 
TCDM ($\Omega_{m}=1.0, \sigma_{8}=0.64, \Gamma=0.25$), 
OCDM ($\Omega_{m}=0.3, \sigma_{8}=1.06, \Gamma=0.25$) and 
LCDM ($\Omega_{m}=0.3, \sigma_{8}=1.22, \Gamma=0.25$). 
In our analysis to derive weak lensing magnification model parameters, we only use the SCDM, OCDM, and LCDM models.
For each simulation, $p(\mu)$ is measured within the redshift range $0.5<z<3.5$. Then we fit Eq.(\ref{eq:P_eta}) to the measurement at each redshift. The best-fit parameters for $\eta_{\mathrm{peak}}$, $\omega$ and $q$ are displayed in Figure \ref{fig:pmu_parameter_fit} as a function of $\xi_{\eta}$. 

The behavior of these quantities show similar dependence on $\xi_{\eta}$, consistent with the universality of $p(\eta)$. 
The parameter $\eta_{\text{peak}}$ is uniform among the three models, but the $\omega$ and $q$ parameters for the LCDM model differ significantly from those from the SCDM and OCDM models. This is
not surprising, since the LCDM model has a much larger $\sigma_8$ than the SCDM and OCDM models. 
Note that the SCDM and OCDM models are tens of $\sigma$ off from the current measurements such as those from Planck \cite{Planck_2018_6},
and the LCDM model has a much higher $\sigma_8$ than current measurements. These three models thus span a much larger parameter space than allowed by current observational data, which means the universality of $p(\eta)$ should be much better than shown in Fig.\ref{fig:pmu_parameter_fit} for viable models. Therefore, we only use the LCDM model, which is closest to current measurements, for deriving the weak lensing magnification model parameters, and do not expect the results to
change significantly for moderate deviations from this LCDM model.

We have performed a polynomial fit of $\eta_{\mathrm{peak}}$, $\omega$ and $q$ as a function of $\xi_{\eta}$, shown as the dashed lines in Fig.\ref{fig:pmu_parameter_fit}.
These fits can be used to derive $p(\eta)$ and $p(\mu)$ for an arbitrary cosmological model at a given redshift, and are as follows:
\begin{eqnarray}
    &&\eta_{peak}=0.4051(\sqrt{\xi_\eta})^{2}-0.6943\sqrt{\xi_\eta}+0.9191 \\
    &&\omega=-0.3231(\sqrt{\xi_\eta})^{2}+0.3867\sqrt{\xi_\eta}+0.3262 \\
    &&q=0.605(\sqrt{\xi_\eta})^{2}-0.5743\sqrt{\xi_\eta}+1.1943
\end{eqnarray}

Given the simulated Roman data set of SNe Ia described in the previous subsection, we add the weak lensing signal by sampling $\mu$ from the probability distribution $p(\mu)$ for each SN Ia, and multiplying its observed flux with $\mu$.
 
\section{Results}
We follow the methodology from our previous analysis, \cite{Zhai_2019}, in reconstructing the weak lensing signal from the realistically simulated Roman data set of SNe Ia described in the previous section. We now summarize the methodology and present the analysis results.

\subsection{Reconstruction of $p(\mu)$}\label{sec:reconstruction}

\begin{figure}%[htbp]
\begin{center}
\includegraphics[width=8cm]{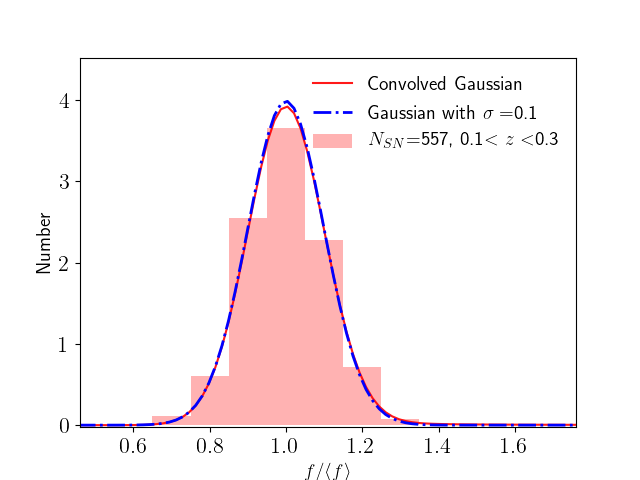}
\caption{The flux distribution of the low-redshift subsample. The redshift range and number of SNe Ia are also shown. The dot-dashed line is a Gaussian distribution with $\sigma=0.1$, obtained from a best-fit algorithm through Eq. (\ref{eq:chi2}). The solid line represents the stretched Gaussian from UPDF for the weak lensing model, which is not significantly different from the Gaussian distribution since the weak lensing effect is marginal at low redshift.}
\label{fig:lowz}
\end{center}
\end{figure}

\begin{figure*}%[htbp]
\begin{center}
\includegraphics[width=18.5cm]{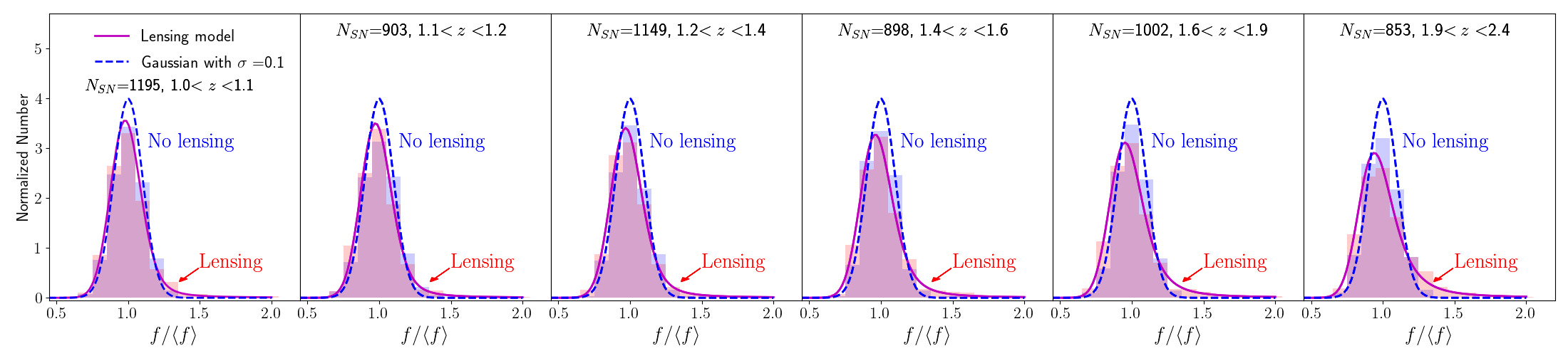}
\caption{The flux distribution for the high-redshift SNe Ia subsamples, split into different redshift ranges. The blue dashed lines represent the intrinsic Gaussian distribution, $g(L_{int})$, with $\sigma$ estimated from the low-redshift subsample. The purple lines represent the convolution of the UPDF model due to weak lensing effect and $g(L_{int})$. The blue histograms show flux distribution when the selection effect is taken into account, while the red histograms show the distribution with weak lensing effect added. Compared with the weak lensing effect, the selection effect is subdominant, and the overall distribution is still consistent with a Gaussian distribution. The weak lensing effect is significantly non-Gaussian, and increases with redshift, and should be properly accounted for in the data analysis.
}
\label{fig:highz}
\end{center}
\end{figure*}

In order to measure the weak lensing signature, we first use the flux-averaging method to find the flux distribution of the SNe Ia as described in \cite{Wang_2005_wl, Zhai_2019}. We use the low-redshift data to anchor the mean flux and the distribution of the intrinsic brightness since the weak lensing effect is negligible at low redshift. The result is presented in Figure \ref{fig:lowz}; the flux distribution at low redshift can be approximated as a Gaussian distribution with $\sigma=0.1$ in units of the mean flux. We split the high-redshift SNe Ia into several bins and the flux distributions are shown in Figure \ref{fig:highz}. We can find similar characteristics as in the real observations \cite{Wang_2005_wl, Zhai_2019}. The observed brightness has a non-Gaussian distribution and the effect increases with redshift.

\begin{figure*}%[htbp]
\begin{center}
\includegraphics[width=18.5cm]{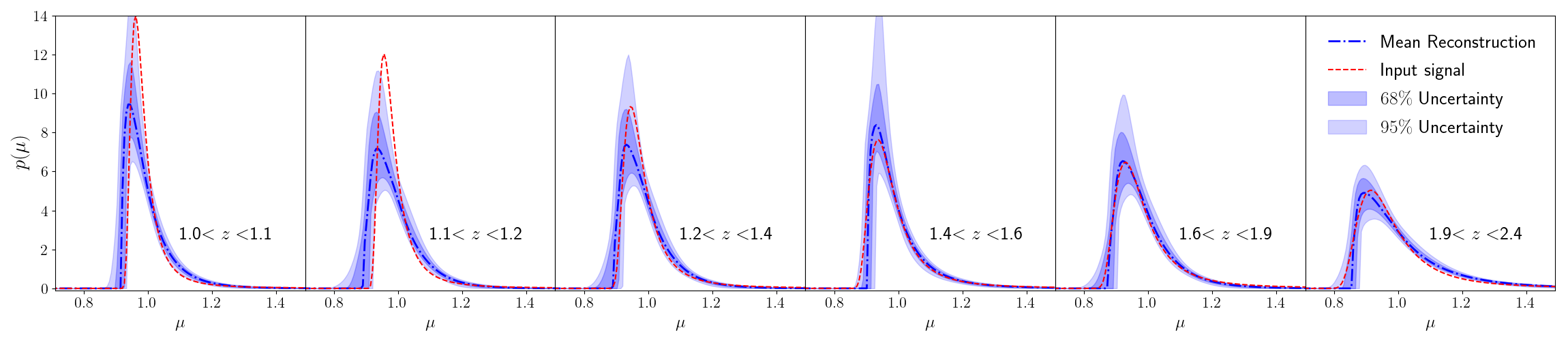}
\caption{Reconstruction of $p(\mu)$ from the high-redshift subsamples with only weak lensing signal added (no other systematics) as shown by the red histogram of Fig.\ref{fig:highz}. The method is based on UPDF as explained in the text. The dot-dashed line represents the mean and the shaded area is the 68\% C.L. range. }
\label{fig:recons}
\end{center}
\end{figure*}

The $p(\mu)$ reconstruction method presented in \cite{Zhai_2019} was based on linear interpolation. In this work, we apply a new method. The flux measurement (Figure \ref{fig:highz}) results from a convolution of intrinsic brightness distribution $g(L_{\text{int}})$ and $p(\mu)$. We assume that $g(L_{\text{int}})$ is independent of redshift and can be derived from the low-redshift observations (Figure \ref{fig:lowz}). In addition, we assume $p(\mu)$ can be described by the stretched Gaussian distribution in Eq.(\ref{eq:P_eta}), and the unknown parameter set $\tilde{P}=\{\mu_{\text{min}}, \eta_{\text{peak}}, \omega, q\}$ can fully determine the observed flux distribution of SNe Ia. Then we adopt the same likelihood as in \cite{Zhai_2019}
\begin{equation} \label{eq:chi2}
    \chi^{2}=\sum_{i=0}^{N_{\text{bin}}}\left(\frac{D_{i,obs}-D_{i,pre}}{\sigma_{D,i}}
    \right)^{2}.
\end{equation}
For a given redshift bin, $D_{i, obs}$ is the number of SNe Ia with flux in the $i-$th bin, $D_{i, pre}$ is the prediction from the lensing model $p(\mu)$, and $\sigma_{D,i}$ is the uncertainty for a Poisson distribution. We estimate the unknown parameter set $\tilde{P}$ through a MCMC analysis using the \texttt{emcee} toolkit \citep{Foreman-Mackey_2013}.

We present the reconstructed result of $p(\mu)$ in Figure \ref{fig:recons} for different redshift bins. It is clear that the reconstruction can capture the main characteristics of the distribution of magnification: a shift of the peak to the faint end due to de-magnification since the universe is mostly empty, and a non-Gaussian tail at the bright end due to high magnifications. It is also clear that this weak lensing signature increases with redshift.

\subsection{Cosmological constraint from weak lensing signature}
\label{sec:lensing_constraints}

\begin{figure}%[htbp]
\begin{center}
\includegraphics[width=8cm]{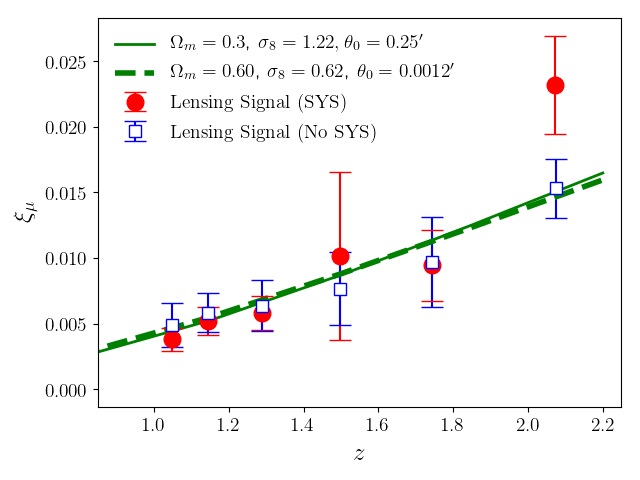}
\caption{Measurements of $\xi_{\mu}$, the variance of the lensing magnification $\mu$ from the simulated SNe Ia data based on our reconstruction method, for both data sets with (SYS) and without systematics (No SYS) added. The lines correspond to different cosmological models and smoothing angle, which can indicate the degeneracy between these parameters.}
\label{fig:xi_mu}
\end{center}
\end{figure}

The MCMC analysis results in the constraints on the parameter set $\tilde{P}$, from which we can estimate the variance of $\mu$. Each model in the MCMC chain accurately describes the shape of $p(\mu)$, but the calculation of $\xi_{\mu}$ needs to be self-consistent. In order to do so, we adopt a similar strategy as in \cite{Wang_2002}: the integral used to calculate the moment, $\int \mbox{d}\mu\, p(\mu)\mu^{2}$, is truncated at $\mu_{\text{max}}$, which is determined by requiring 
\begin{equation}
\langle\mu\rangle \equiv \int_0^{\mu_{\text{max}}} \mbox{d}\mu\, p(\mu)\mu=1. 
\end{equation}
This prevents the contribution of the noisy high-$\mu$ tail from impacting the mean significantly. For the reconstruction at each redshift bin, we adopt this method and obtain the measurements of $\xi_{\mu}$ as shown in Figure \ref{fig:xi_mu}. Two different sets of measurements are shown in Figure \ref{fig:xi_mu}: the "SYS" set consists of the $\xi_{\mu}$ measured from the simulated Roman data set of SNe Ia with systematic effects as discussed in Sec.\ref{sec:sys}, while the "No SYS" set consists of $\xi_{\mu}$ measured from a reference set of SNe Ia with the same redshift and distance modulus uncertainty for each SN Ia, but with its distance modulus replaced by the prediction from the true cosmological model (i.e., the input model for the SN Ia data simulation). We summarize these different situations in Table \ref{tab:terms}. 

\begin{table}
\centering
\begin{tabular}{c|l}
\hline
term &  characteristics \\
\hline
SYS & Realistic SN sample \\
no SYS & Ideal SN sample \\
weak lensing & add lensing signal to SN Ia brightness \\
\hline
\end{tabular}
\caption{A summary of the different cases considered in this analysis.}
\label{tab:terms}
\end{table}

For a given cosmological model and redshift, the theoretical prediction for $\xi_{\mu}$ can be calculated from Eq. (\ref{eq:xi_mu}). We use the transfer function from \cite{Eisenstein_1998} to calculate the matter power spectrum for simplicity and the halofit model to add the non-linear correction \cite{Smith_2003, Takahashi_2012}. In addition, in the framework presented in Section \ref{sec:signature}, one still needs to determine $\theta_{0}$, the smoothing angle. This parameter sets the scale of the window function and cuts the power at small scale, which however cannot be ignored for point sources like SNe Ia \cite{Valageas_2000a}. Figure \ref{fig:xi_mu} shows the predicted $\xi_{\mu}$ for two different cosmological models. Clearly, $\Omega_{m}$ and $\sigma_{8}$ are degenerate, as expected, as a common feature in weak lensing analysis, e.g. \cite{Kohlinger_2017}. In addition, the value of the smoothing angle also impacts the overall amplitude of the lensing signal, which is degenerate with the cosmological parameters as well.

\begin{figure*}%[htbp]
\begin{center}
\includegraphics[width=8.5cm]{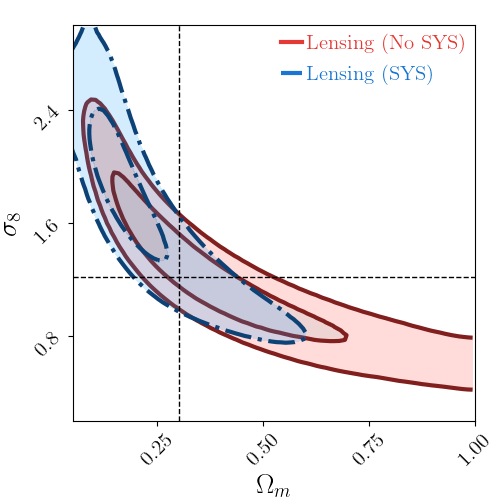}
\includegraphics[width=8.5cm]{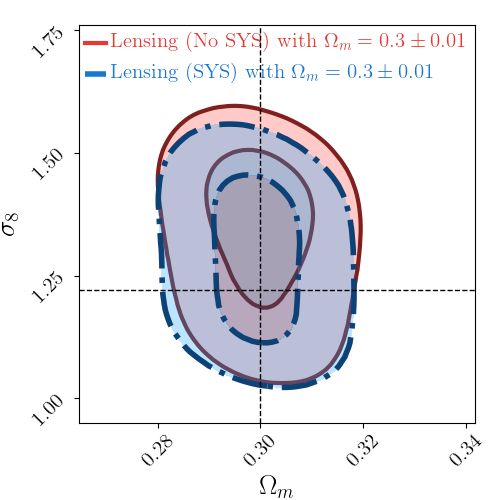}
\caption{Cosmological constraint from the measurement of $\xi_{\mu}$ with the simulated SNe Ia data. Both cosmological parameters and the smoothing angle are allowed to vary as free parameters. $Left~Panel:$ Flat prior for all the parameters. $Right~Panel:$ prior of $\Omega_{m}=0.3\pm0.01$ is applied. The contours in the figures show 1 and 2$\sigma$ confidence levels.}
\label{fig:cons}
\end{center}
\end{figure*}

We have carried out an MCMC likelihood analysis, comparing the measured and predicted $\xi_{\mu}$ values at various redshifts, to derive robust cosmological constraints.  Figure \ref{fig:cons} shows the resultant joint confidence contours on $\Omega_m$ and $\sigma_8$, marginalized over the smoothing angle $\theta_{0}$ with the flat prior $0<\theta_{0}<1.0$. The left panel shows the results from SN Ia weak lensing magnification only. The right panel shows the results of SN Ia lensing magnification with a prior on $\Omega_m$, $\Omega_m=0.3\pm 0.01$.
We find that the input cosmology can be recovered within $1\sigma$ using only SN Ia lensing data without systematic effects. When the systematic effects are included, the parameter estimates are biased by more than $1\sigma$, unless a prior on $\Omega_m$ is added. This kind of test on simulated SN Ia data can be used to identify and mitigate systematic effects.

The distance measurements from Roman SNe Ia alone provide a powerful probe of cosmic expansion history. We use flux-averaging to remove/minimize the effect of weak lensing magnification of SNe Ia in analyzing SN Ia distance measurements \citep{Wang_2000}.
In the simplest model of a flat universe with a cosmological constant,
we obtain $\Omega_{m}=0.303\pm0.003$ (SYS) and $\Omega_{m}=0.298\pm0.003$ (No SYS) respectively, using the distance measurements from the simulated Roman data set of SNe Ia. Expanding the cosmological model to include more parameters (to be investigated in future work) will lead to significantly larger uncertainty on the $\Omega_m$ measurement. We use $\Omega_{m}=0.3\pm 0.01$ as a proxy of such an analysis of SN Ia distance measurements, and the $\Omega_m$ prior on the SN Ia lensing data, to illustrate the power of combining SN Ia lensing and distance measurements, see the right panel of Figure \ref{fig:cons}. Note that the addition of the $\Omega_m$ prior removes the bias in the
estimated parameters in the presence of systematic effects, and tightens the constraints on $\sigma_8$. 

\begin{figure*}%[htbp]
\begin{center}
\includegraphics[width=8.5cm]{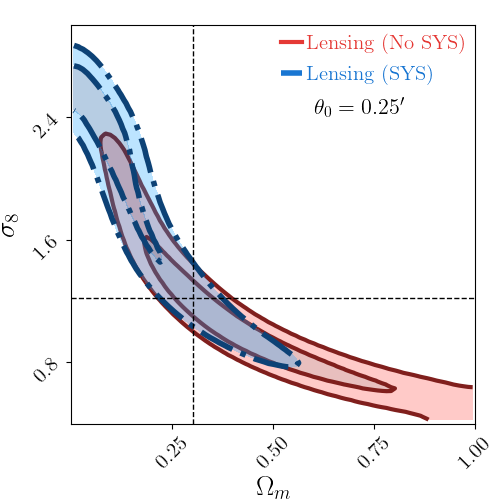}
\includegraphics[width=8.5cm]{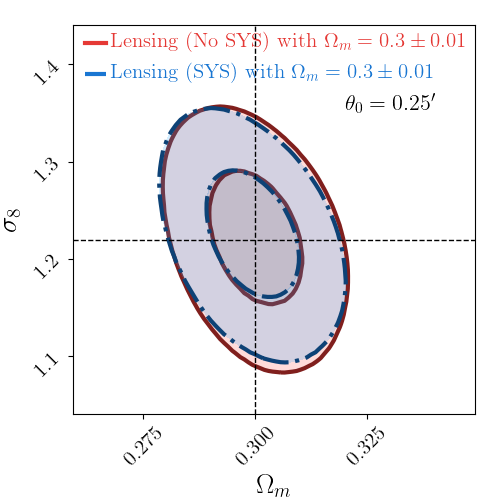}
\caption{Cosmological constraint from the measurement of $\xi_{\mu}$ with the simulated SNe Ia data, with fixed smoothing angle $\theta_{0}=0.25'$. $Left~Panel:$ Flat prior for all the parameters. $Right~Panel:$ prior of $\Omega_{m}=0.3\pm0.01$ is applied. The contours in the figures show 1 and 2$\sigma$ confidence levels.}
\label{fig:cons2}
\end{center}
\end{figure*}

$\Omega_m$ and $\sigma_8$ are degenerate with the smoothing angle $\theta_0$. This is illustrated in Figure \ref{fig:xi_mu}, where the $\xi_\mu$ measurements without systematics can be fit equally well by two very different cosmological models with different values of $\theta_0$.
The information on $\theta_0$ is the key to further tighten the cosmological constraints.
In principle, $\theta_0$ can be determined from ray-tracing simulations based on cosmological N-body simulations for different cosmological models. 
Figure \ref{fig:xi_mu} shows that given a cosmological model, the $\xi_\mu$ measurements can be fitted to determine $\theta_0$, which is found to be 0.25$^\prime$ for our assumed true cosmological model. Fig.\ref{fig:cons2} shows the $\Omega_m$ and $\sigma_8$ joint confidence contours, with the same line types as in Fig.\ref{fig:cons}. As expected, fixing $\theta_0$ significantly tightens cosmological constraints.

The constraints on $\sigma_{8}$ are summarized in Table \ref{tab:cons}. We find that $\sigma_8$ can be msaured to $\sim$ 10\% using SNe Ia data alone, and $\sim$ 5\% if the smoothing angle $\theta_0$ can be determined from ray-shooting simulations.

\begin{table}
\centering
\begin{tabular}{c|ll}
\hline
\multicolumn{3}{c}{ Input model: $\Omega_{m}=0.3, \sigma_{8}=1.22$ }    \\
\hline
 &  Data used              & $\sigma_{8}$   \\
\hline
free $\theta_{0}$ & $\xi_{\mu}$ (no SYS)   &   $1.101_{-0.369}^{+0.586}$  \\
 & $\xi_{\mu}$ (SYS) & $1.694_{-0.601}^{+0.532}$  \\
 & $\Omega_{m}=0.3\pm0.01~+\xi_{\mu}$ (no SYS)   & $1.340_{-0.169}^{+0.131}$  \\
 & $\Omega_{m}=0.3\pm0.01~+\xi_{\mu}$ (SYS) & $1.289_{-0.147}^{+0.145}$  \\
\hline
$\theta_{0}=0.25^\prime$ & $\xi_{\mu}$ (no SYS)     & $0.997_{-0.349}^{+0.581}$ \\
& $\xi_{\mu}$ (SYS)  & $1.922^{+0.540}_{-0.761}$ \\
& $\Omega_{m}=0.3\pm0.01~+\xi_{\mu}$ (no SYS)   & $1.221_{-0.066}^{+0.064}$  \\
& $\Omega_{m}=0.3\pm0.01~+\xi_{\mu}$ (SYS) & $1.224_{-0.062}^{+0.062}$  \\
\hline
\end{tabular}
\caption{Constraints on $\sigma_{8}$ with SNe lensing signal. The uncertainties represent the centering 68\% distribution around the point with peak probability. The bottom section shows constraint by fixing $\theta_{0}=0.25'$.}
\label{tab:cons}
\end{table}

\section{Summary and Discussion}

We have presented a methodology for reconstructing the probability distribution of the weak lensing magnification $\mu$ of SNe Ia $p(\mu)$ from observational data, and using it to constrain cosmological parameters, and applied it to simulated Roman data set of SNe Ia. We find that using the weak lensing magnification of the SNe Ia constrains a combination of matter density $\Omega_m$ and matter clustering amplitude $\sigma_8$. SN Ia distances alone lead to a better than 1\% measurement of $\Omega_m$. The combination of SN Ia weak lensing magnification and distance measurements result in a $\sim$ 10\% measurement on $\sigma_8$. The SNe Ia from the Roman Space Telescope will be powerful in constraining the cosmological model.

This work extends our earlier paper for the reconstruction of the weak lensing magnification distribution from SNe Ia observations \cite{Zhai_2019}. We have revisited the universality of the function $p(\eta)$ to model the weak lensing magnification of SNe Ia, and derived new fitting formulae for calculating $p(\mu)$ as a stretched Gaussian in an arbitrary cosmological model. 
Using the realistically simulated Roman data set containing 14,000 SNe Ia, 
we have successfully reconstructed $p(\mu)$ in an MCMC analysis. We find that for a redshift bin at $z>1.0$, a few hundreds of SNe Ia can form a statistically sufficient sample to enable useful reconstruction of $p(\mu)$. 

In another MCMC likelihood analysis comparing the variance of $\mu$ measured from the reconstructed $p(\mu)$ to its theoretical prediction, we find that SN Ia lensing magnification constrains a combination of $\Omega_m$ and $\sigma$, as expected for weak lensing measurements, but the $\Omega_m$ constraint from SN distance measurements breaks that degeneracy and leads to tight constraints on $\sigma_8$. We find that both  $\Omega_m$ and $\sigma$ are degenerate with the smoothing angle $\theta_0$, a paramter introduced in the modeling of weak lensing magnification, which could in principle be determined via ray-tracing experiments on cosmological N-body simulations. The information on $\theta_0$ leads to the tightest constraints on $\sigma_8$.

The measurement of $\xi_{\mu}$ from the weak lensing magnification of SNe Ia provides an independent cosmological probe, complementary to the SN Ia distance modulus. This observable is worth further investigation in the future from the aspects of both theoretical modeling and observational analysis. 

We note that the reconstruction method adopted in this work is not unique, and other parameteric or non-parameteric method is also possible. However, the reconstructed result and the measurement of the moments of $\mu$ shouldn't change significantly. On the other hand, only the variance of $\mu$ is used in our analysis. This means that adding the information from higher order moments may have more constraining power and have different parameter dependence. A method like ``MeMo" \cite{Marra_2013, Quartin_2014} can be useful in this investigation and this also requires precise calibration based on numerical simulations and thus we will leave this for future work. 

The analysis presented in this work demonstrates that the SNe Ia can be used not only as a geometrical probe of cosmic expansion, but also a probe of the clustering of matter in the universe. Our results indicate that Roman data set of SNe Ia will place powerful constraints on the cosmological model.

\section*{Acknowledgements}

This work was supported in part by NASA grant 15-WFIRST15-0008 Cosmology with the High Latitude Survey Roman Science Investigation Team (SIT). DS is supported in part by NASA under Contract No. NNG17PX03C issued through the WFIRST Science Investigation Teams Programme.  DS is also supported by DOE grant DE-SC0010007 and the David and Lucile Packard Foundation.

\bibliography{DE_GoF,software}

\end{document}